\documentclass[a4paper,11pt]{article}
\usepackage{authblk}
\usepackage{txfonts}
\usepackage{graphicx}
\usepackage[a4paper,left=3cm,right=2cm,top=2.5cm,bottom=2.5cm]{geometry}

\newif\ifnatbibsort\natbibsorttrue
\ifnatbibsort\usepackage[numbers,sort&compress]{natbib}\else\RequirePackage[numbers,compress]{natbib}\fi

\usepackage[colorlinks=true
,urlcolor=blue
,anchorcolor=blue
,citecolor=blue
,filecolor=blue
,linkcolor=blue
,menucolor=blue
,pagecolor=blue
,linktocpage=true
,pdfa=true
]{hyperref}

\pdfoutput=1 

\title{\boldmath First functionality tests of a 64 x 64 pixel DSSC sensor module connected to the complete ladder readout}

\author[a,1]{M. Donato,\footnote{Corresponding author. E-mail: mattia.donato@xfel.eu}}
\author[b]{K. Hansen,}
\author[b]{P. Kalavakuru,}
\author[c]{M. Kirchgessner,}
\author[a]{M. Kuster,}
\author[a]{M. Porro,}
\author[b]{\mbox{C. Reckleben}}
\author[a]{and M. Turcato}

\affil[a]{European XFEL GmbH,
	Holzkoppel 4,
	Schenefeld, Germany}

\affil[b]{Deutsches Elektronen-Synchrotron DESY,
	Notkestra\ss e 85,
	Hamburg, Germany}

\affil[c]{Heidelberg University, ZITI, 
	B6,23-29, 68159 Mannheim, Germany}

\begin{document}
\maketitle
Prepared for submission to JINST\\
18$^{\rm{th}}$ International Workshop on Radiation Imaging Detectors,
3$^{\rm{rd}}$ - 7$^{\rm{th}}$ July 2016,
Barcelona, Spain\\

\begin{abstract}
The European X-ray Free Electron Laser (XFEL.EU) will provide every 0.1 s a train of 2700 spatially coherent ultrashort X-ray pulses at $4.5$ MHz repetition rate. The Small Quantum Systems (SQS) instrument and the Spectroscopy and Coherent Scattering instrument (SCS) operate with soft X-rays between $0.5$ keV - $6$ keV.\\
The DEPFET Sensor with Signal Compression (DSSC) detector is being developed to meet the requirements set by these two XFEL.EU instruments. The DSSC imager is a 1 mega-pixel camera able to store up to 800 single-pulse images per train.
The so-called ladder is the basic unit of the DSSC detector. It is the single unit out of sixteen identical-units composing the DSSC-megapixel camera, containing all representative electronic components of the full-size system and allows testing the full electronic chain. Each DSSC ladder has a focal plane sensor with $128\times 512$ pixels.
The read-out ASIC provides full-parallel readout of the sensor pixels. Every read-out channel contains an amplifier and an analog filter, an up-to 9 bit ADC and the digital memory. The ASIC amplifier have a double front-end to allow one to use either DEPFET sensors or Mini-SDD sensors. In the first case, the signal compression is a characteristic intrinsic of the sensor; in the second case, the compression is implemented at the first amplification stage. The goal of signal compression is to meet the requirement of single-photon detection capability and wide dynamic range.\\
We present the first results of measurements obtained using a $64\times 64$ pixel DEPFET sensor attached to the full final electronic and data-acquisition chain. 	
\end{abstract}

\flushbottom
\newpage

\section{Introduction}
\label{sec:intro}

\begin{figure}[htbp]
	\centering 
	\includegraphics[width=.45\textwidth,clip]{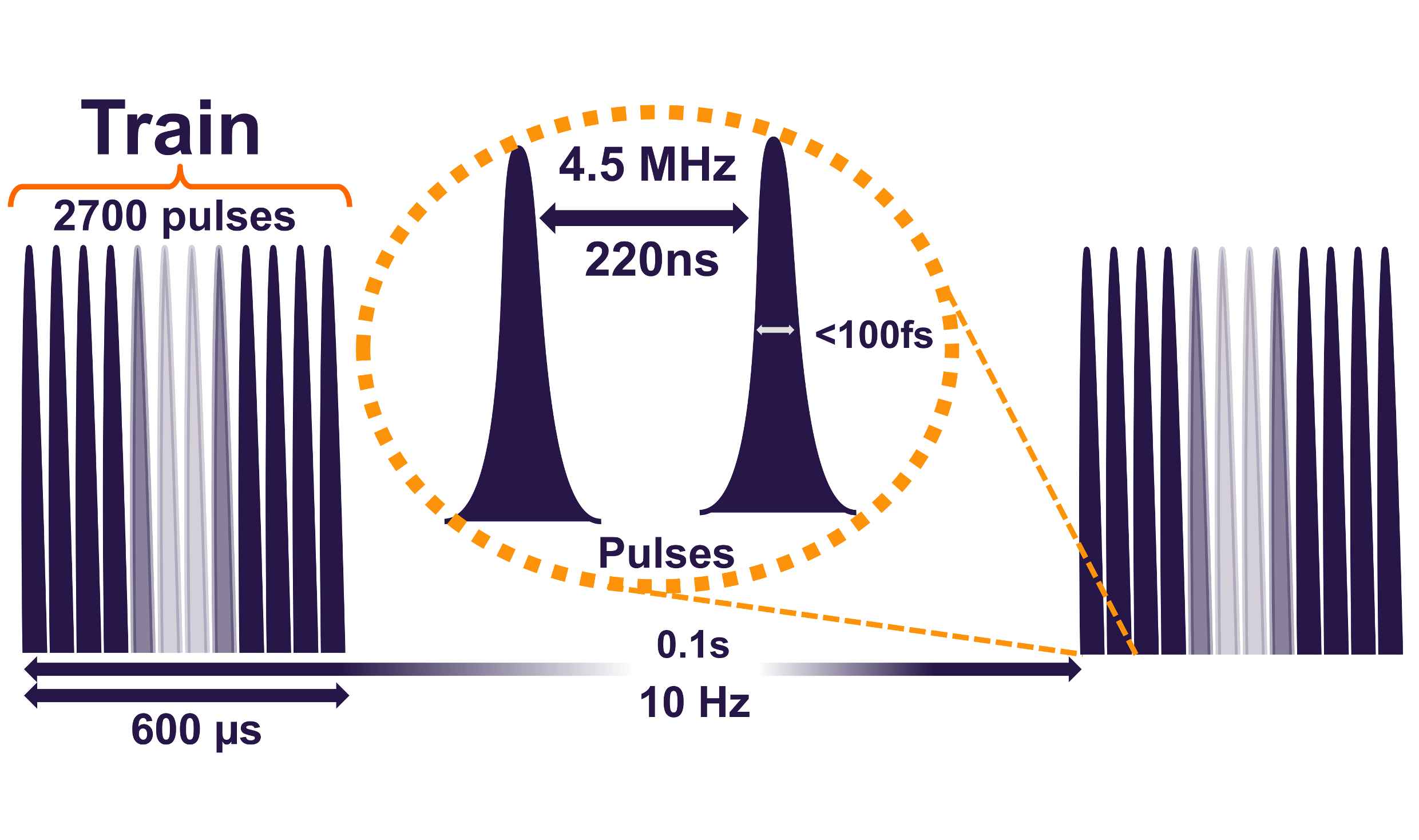}
	\qquad
	\centering 
	\includegraphics[width=.45\textwidth,clip]{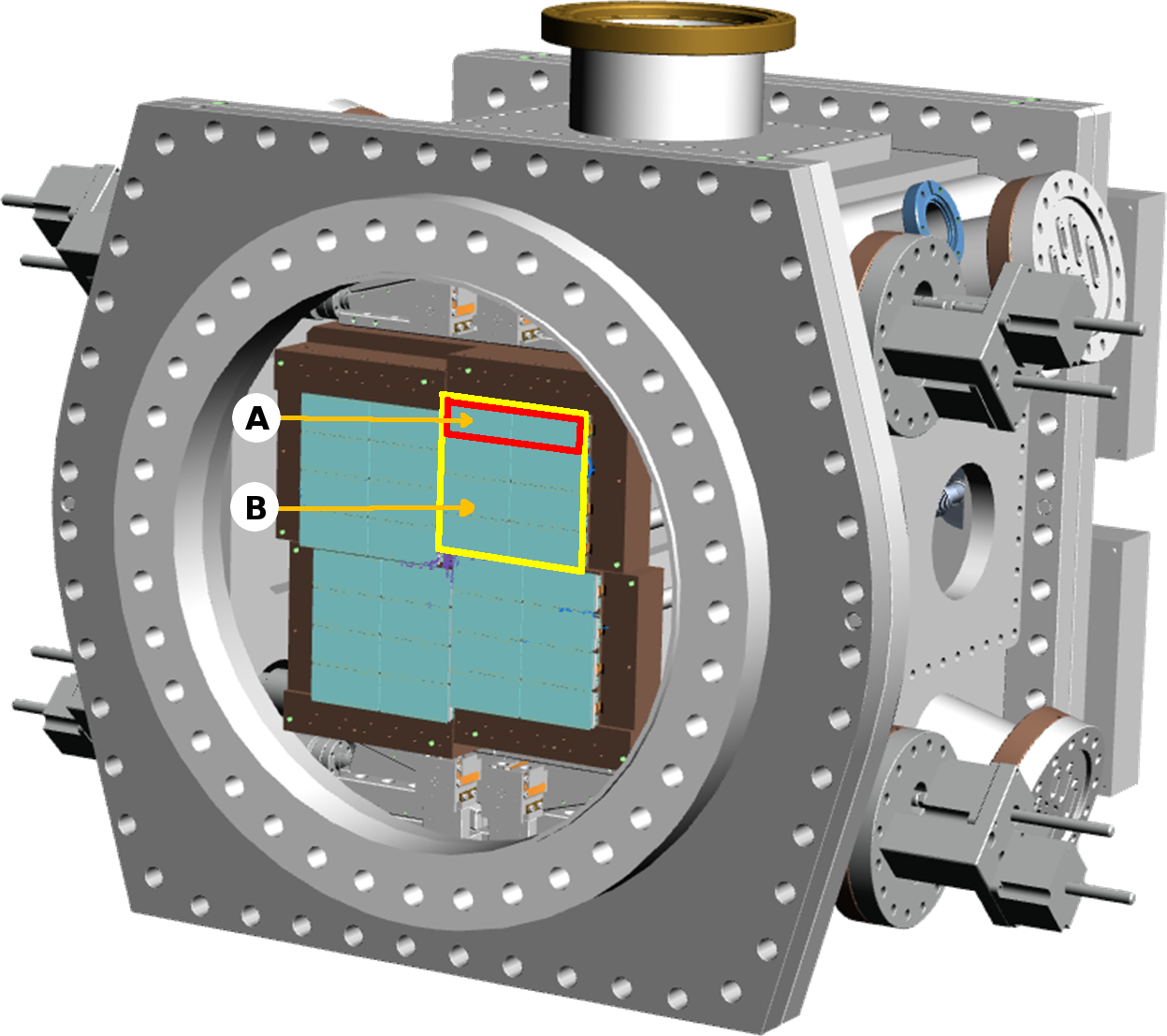}
	\caption{\label{fig:time} On the left, the XFEL.EU time structure. On the right, a 3D CAD model of the DSSC megapixel camera and its vacuum vessel is shown. \label{fig:megapixel} The label A indicates the DSSC ladder and the label B indicates the quadrant.}
\end{figure}

The European X-ray Free Electron Laser (XFEL.EU) will provide ultra-short spatially-coherent X-rays pulses with a unique \mbox{time-structure \cite{xfelreport}}. As shown in Figure \ref{fig:time}, a train of $2700$ X-ray pulses with $220$ ns separation time are repeated every $0.1$ s. 
At XFEL.EU, the Small Quantum Systems (SQS) instrument and the Spectroscopy and Coherent Scattering instrument (SCS) operate with soft X-rays between $0.5$ keV - $6$ keV. The SCS instrument aims at the investigation of electronic and atomic structure and dynamics of soft matter, biological structures, magnetic materials and structures \cite{SCS}. The SQS instrument will investigate processes in atoms, ions, small molecules and clusters occurring under highly intense beams using a variety of spectroscopy techniques \cite{SQS}. \\

The DEPFET Sensor with Signal Compression (DSSC) detector \cite{matteo} is being developed to meet the requirements set by the SCS and SQS instruments. Due to the requirements of the soft-X-rays beamlines, the detector will be operated under high vacuum conditions.

The DSSC imager will be a \mbox{1 Mpixel} camera able to store up to 800 single-pulse images per XFEL.EU train matching the XFEL.EU time structure. It is optimized for the energy range $0.5~{\rm keV} - 6~{\rm keV}$ and with the possibility to detect single photons in each frame.  The pixel shape is hexagonal, with a pixel size of $\sim 204~{\rm\mu m} \times 236~{\rm \mu m}$. The single ASIC supports $64 \times 64$ pixels and the readout provide per-pixel amplification and filtering, an ADC and a digital memory. At standard XFEL.EU pulse repetition rate ($4.5 {\rm MHz}$) the ADC has 8 bit of resolution and at slower speed ($\le 2.2 {\rm MHz}$) an extra bit can be used to extend the ADC range. To meet the requirement of single-photon detection capability and wide dynamic range ($\sim 10^4$ photons/pixel/pulse), the detector has been designed to provide a response with a defined compression curve \cite{matteo}. The readout ASIC \cite{asic} \cite{fiorini} implements two selectable front-ends. They allow the system to be operated either with DEPFET or MiniSDD pixel arrays. In the first case, the signal compression is a characteristic of the sensor; in the second case, the compression is implemented electronically in the ASIC front-end. 

\subsection{The DSSC ladder camera}
\label{ladd}
The DSSC megapixel camera \cite{karsten} is composed of four quadrants, see Figure \ref{fig:megapixel} (right). Every quadrant comprise four ladders of the size of $512 \times 128$ pixels. From the electronics point-of-view, a single DSSC ladder is an independent unit and it contains all representative components of the full-size system.

\begin{figure}[htbp]
	\centering 
	\includegraphics[width=1\textwidth,clip]{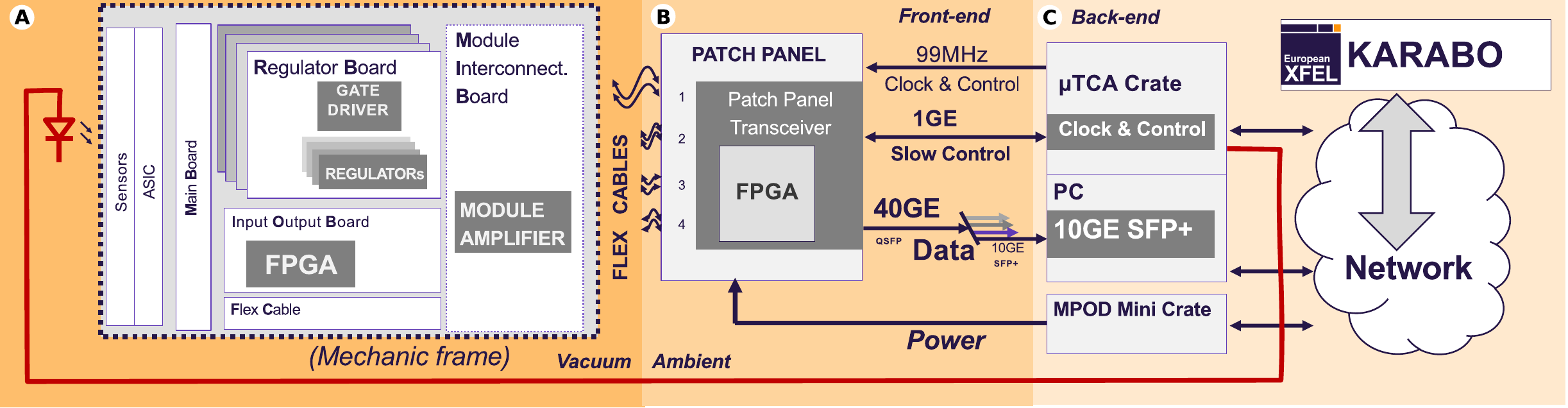}
	\caption{\label{fig:schemedssc} Frame A shows the components placed in vacuum, the mechanic frame including the electronics and the flex-cables. Frame B shows the patch panel and its links. These last devices are located outside vacuum. The interface between vacuum (A) and ambient (B) is given by a flex-cable glued to a feed-through flange. Frame C shows all the interfaces to the external components.
	The red line represents the connection used to synchronize an external LED for test purposes.}
\end{figure}

\begin{figure}[htbp]
	\centering 
	\includegraphics[width=1\textwidth,clip]{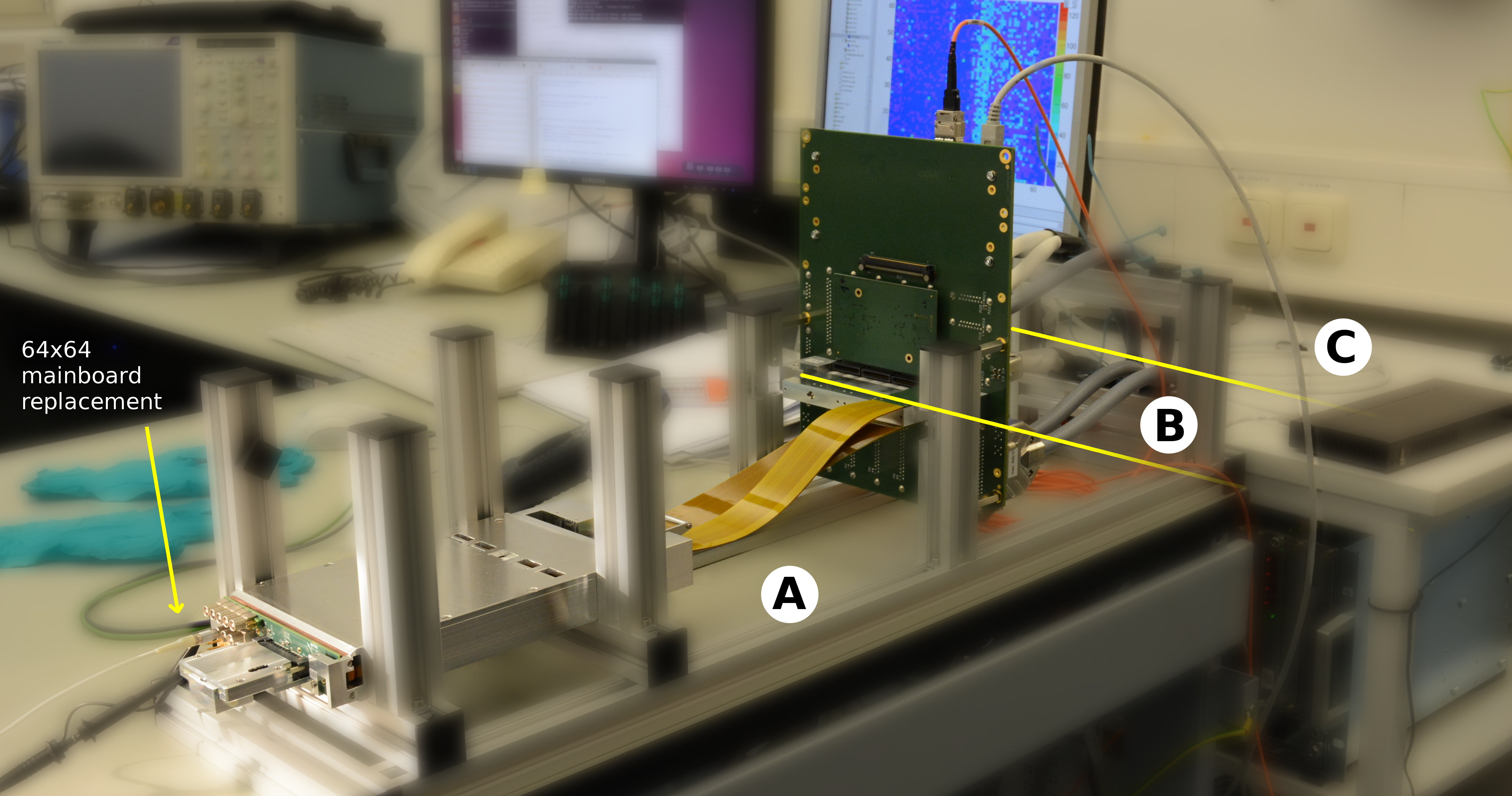}
	\caption{\label{fig:photoladder} DSSC ladder in clean room, equipped with the single ASIC.
		The zone A includes the components that can stay in vacuum such as the mechanic frame of the ladder holding sensors, ASICS and other components. The zone B corresponds to the patch panel and its links and they will stay in ambient environment. The zone C includes the power supplies and the external components. }
\end{figure}

A block scheme of the DSSC ladder is shown Figure \ref{fig:schemedssc}. The focal-plane, comprising the sensors, the ASICs and the main board is connected to four regulator boards, to the input/output board (IOB) and to the flex cable, connected together to the module interconnection board. These boards are placed in vacuum;  they are connected via a flex cable to the patch panel, which is placed in ambient, outside of the DSSC vessel. The patch panel is attached to a vacuum feedthrough, providing an appropriate connection for the patch panel flex cable.

The patch panel provides the sockets for the power-supply cables and a slot for the patch panel transceiver (PPT). The PPT \cite{manfred} is a FPGA-based board (Xilinx Vertex 6), which programs the pixel register of the ASICs, controls the IOBs, synchronizes them to the reference standard XFEL Clock and Control (CC) \cite{cc} signals (such as master clock, start/stop, bunch id, veto); it also converts the data in UDP network packets according to the standard XFEL train data format \cite{XTDF} and transmits them through the optic link. The PPT therefore provides the connections to the slow control system, to the standard CC device, and to the optic fiber for data transfer. The type of optic transceiver used for the data is a Quad Small Form-factor Pluggable (QSFP+) which corresponds to four independent $10$ Gbit/s Ethernet (GE) optic lines. A fiber optic breakout cable (40GE QSFP+ to 4 $\times$ 10GE SFP+) is used to split the four independent lines and each line is assigned to a single ladder. In case of the standard XFEL timing, the average data throughput for a single ladder  is $\sim 1$ GByte/s. In the ladder test system, the data from a single line are acquired by a fast PC equipped with a 10GE PCIe board (Intel 82599ES). The software data receiver is buffering on the RAM consecutive bursts and writes them on two fast Solid State Disks (SSD) in \mbox{RAID $0$} configuration cyclically. In the final megapixel camera the data acquisition will be managed by the XFEL.EU Train Builder \cite{trainbuilder} and XFEL.EU PC Layer \cite{datarecording}, that will be able to store and manage all bursts. The detector control, data acquisition and data processing is integrated into the XFEL.EU control framework Karabo \cite{karabo}.

The regulator boards are local power-supplies used for the power-cycling of the ASICs. In the case of the DEPFET sensors, the gate driver sitting on the regulator boards generates the strong current pulses used to remove the charges from the internal gate of the sensors; they are therefore needed in order to reset the sensors. In case of Mini-SDD sensors, the gate driver is not used.
The IOB is an FPGA-based board (Xilinx Kintex 6) used for generating the timing signals to manage the RBs power-cycling and switching the sensors power lines; it also controls and reads out the data from the ASICs and transmits them to the PPT.
The only active device on the module interconnection board is the module amplifier used to pre-amplify the timing signal from the IOB to the regulator boards. The JTAG line (used for programming the ASICs) and the high voltages (needed for the sensors) are connected via the Flex Cable (FC). 

\section{Measurement}
In this paper, measurements performed using the full read-out chain are presented for the first time. First functionality tests have been performed using a LED. This has the advantage of providing a pulsed source which can be synchronized to the detector in a simple and efficient way.

\subsection{Description of the measurement}

The first measurements performed with the full DSSC ladder read-out chain have been performed under ambient conditions with a single ASIC board, which replaces the 16 ASICs mainboard. For these tests a DEPFET sensor prototype with $128 \times 64$ pixel equipped with a single $64\times64$ pixel ASIC has been used. From the electronic point of view, there are no major differences with respect to the final readout-chain as described in the first section, except that only a single ASIC is used.
\begin{figure}[htbp]
	\centering 
	\includegraphics[height=.45\textwidth,clip]{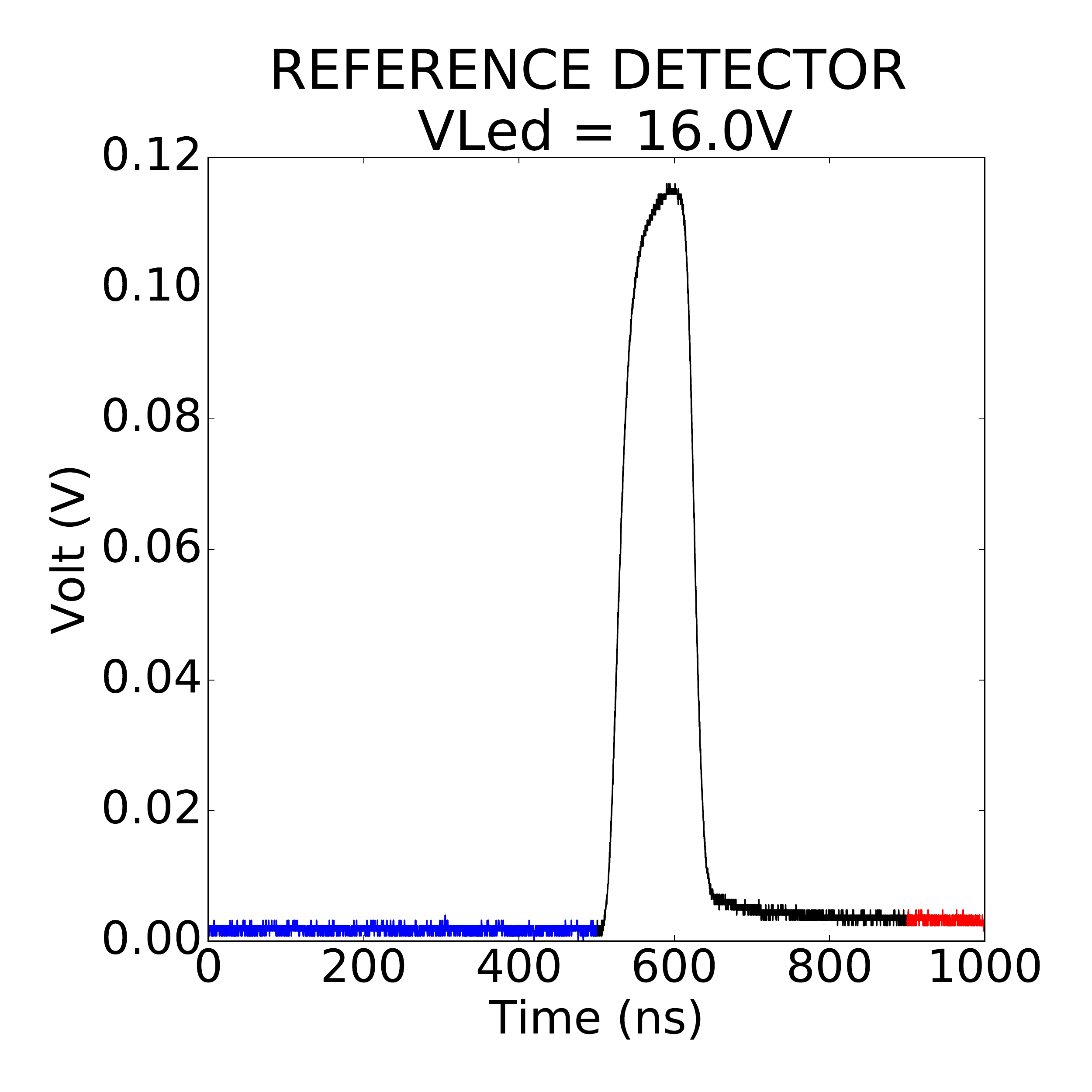}
	\qquad
	\centering 
	\includegraphics[height=.45\textwidth,clip]{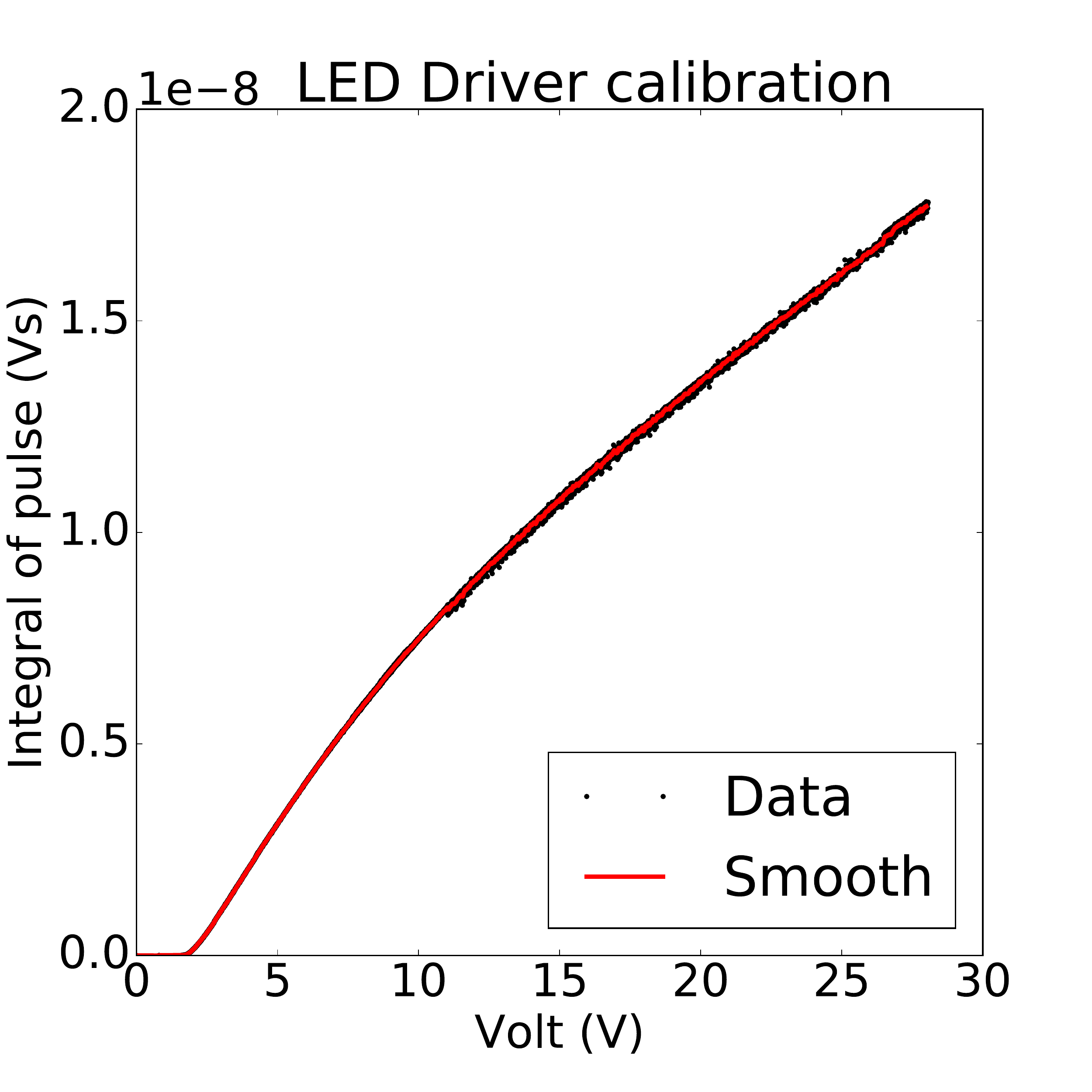}
	\caption{\label{fig:ledcurve} On the left: the LED-light pulse seen by the reference photodiode, measured with the oscilloscope is shown. On the right: the relation between the input voltage of the LED driver vs integrated pulse is shown.}
\end{figure}
\label{sec:LEC}
In order to generate light pulses synchronized to the detector readout cycle, the DSSC ladder electronics has been connected to the CC. This board, used in stand-alone mode, is able to generate the standard XFEL $99~{\rm MHz}$ master clock and emulates the start/stop train signal with a cycle duration of $0.1$ s. Furthermore, it can generate a defined pulse with a fixed delay from the start train signal. Figure \ref{fig:schemedssc} shows a simplified scheme of the connection. 
The LED driver triggered by the CC sends one light pulse every bunch train, emulating the time structure of XFEL.EU.

The LED driver is composed of three stages: an LVDS to TTL signal stage (FIN1002) and a TTL amplifier composed by multiple buffer (AC541) and a fast MOSFET (FDG8850NZ) that can drive \mbox{up to $\sim 2 {\rm A}$}. The current flowing in the LED -- and therefore the light intensity -- is given by the voltage applied to the LED driver, where this voltage is provided by a programmable power supply. The LED used is a red LED (624 nm), $5$ mm large, the max current nominally supported is $20$ mA in continuous mode. This last value can become more than 10 times higher in single-pulse mode. The LED has been set to a fixed duration pulse, defined by the CC signal, and the LED power-supply voltage (VLed) has been scanned from $0$ V $-$ $29$ V.

The light intensity depends on the applied voltage Vled and has been calibrated by using the reference detector, an amplified Si photodiode from ThorLabs (DET10A). 
The left side of \mbox{Figure \ref{fig:ledcurve}} shows a pulse seen by the photodiode. The difference between different LED pulses at a fixed VLed is negligible compared to the electronic noise seen by the oscilloscope. The blue part is considered as baseline, the black part is the part of the signal seen by the detector ($\sim 400{\rm ns}$) and the red part is the tail generated by electronic reflections in the cable between the LED and its driver. \\ 
On the right side, the relation between the input voltage of the LED driver vs integrated pulse is shown. \\

\subsection{Synchronization}
In this first test, the DSSC ladder electronics with single ASIC board has been set to operate with a sequencer cycle length of 100 which corresponds to $100~/~99~{\rm MHz}=1010~{\rm ns}$, equivalent to a frame rate of $\sim 1{\rm MHz}$. The detector has been configured to use the ADC in 8 bit mode and to store sequential frames in sequential memory cell. The charge-collection time window has been set to $403~{\rm ns}$ and the reset is $36~{\rm ns}$ long. 
\begin{figure}[htbp!!!]
	\centering 
	\includegraphics[width=.98\textwidth,clip]{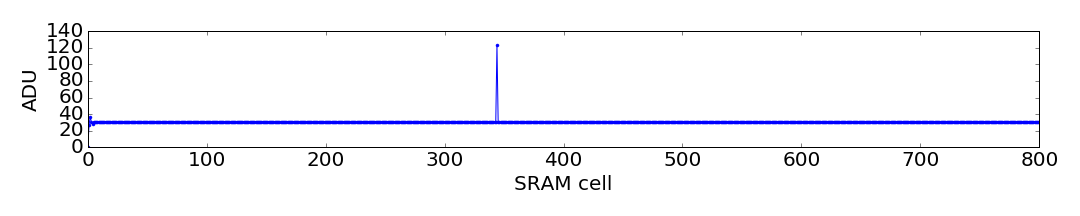}
	\caption{\label{fig:singlepulse} The memory cell content of a single pixel of the DSSC ladder during the illumination with the LED pulse.}
\end{figure}

Figure \ref{fig:singlepulse} shows the content of 800 memory cells of a single illuminated pixel and it shows that the light pulse is observed in a single memory cell. By tuning the CC it is possible to move the timing of the LED signal with a granularity of $\sim 10~{\rm ns}$. By setting different increasing time delays, we observe lower (down to the baseline) signals on the detector when the light pulse is not fully inside (and eventually falls outside) the detector integration window. The exact synchronization between the LED and the detector has been determined by identifying the point in time when the intensity of the signal on the detector is maximal. By repeating this test, as expected, the light pulse is observed always in a single and fixed memory cell, and with a stable content. This first result shows that the detector is correctly synchronized to the CC signals.

\begin{figure}[htbp!!!]
	\centering
	\includegraphics[height=.34\textwidth,clip]{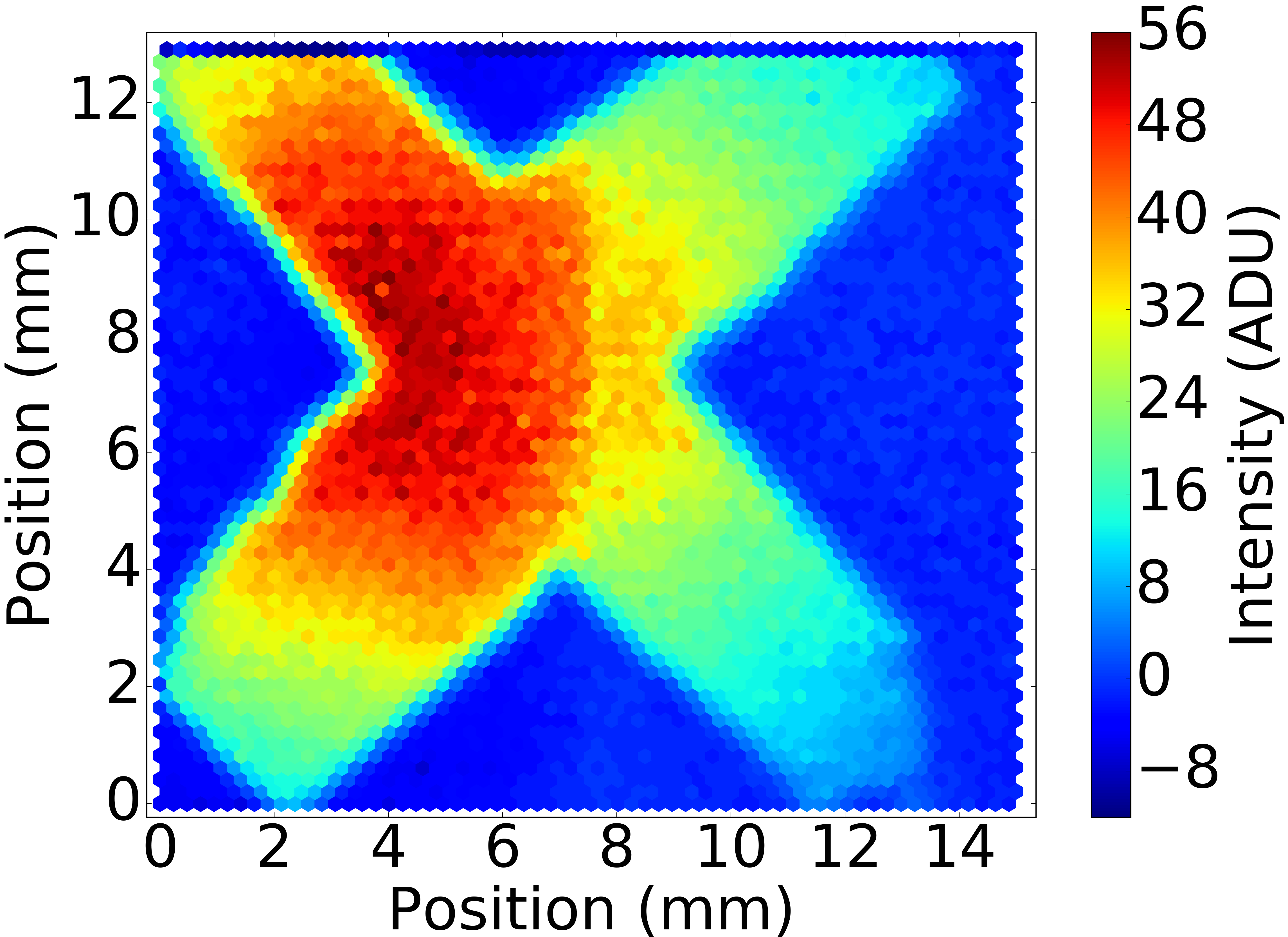}	
	\qquad
	\centering
	\includegraphics[height=.34\textwidth,clip]{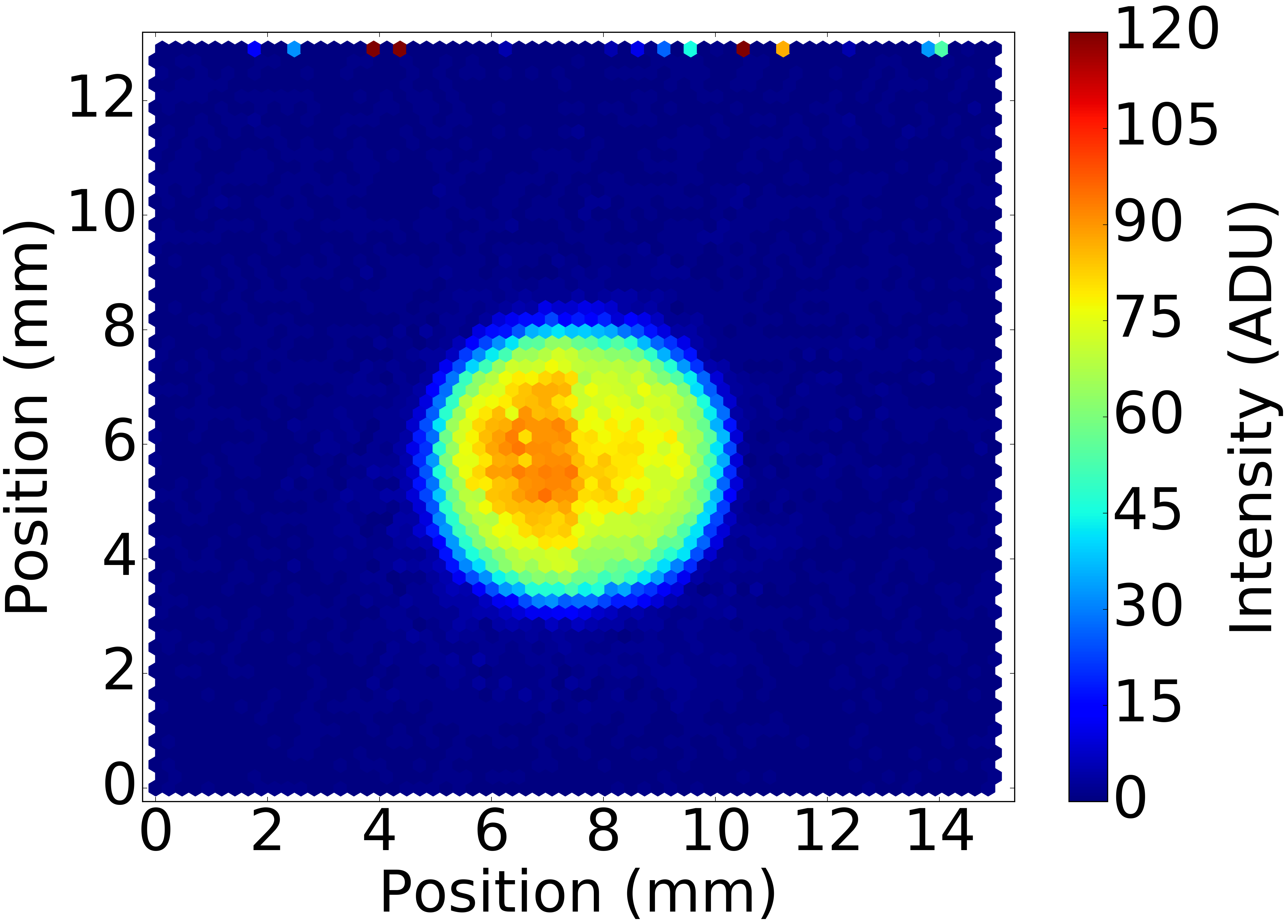}	
	\caption{\label{fig:LEDimgLIGHT} On the right an image of an illuminated paper-mask located between the detector and the light source is shown. On the left an image of the directly illuminated sensor is shown. The LED is located at a distance of a few millimeters from the sensor.}
\end{figure}

Figure \ref{fig:LEDimgLIGHT} (right) shows a single image collected by the DSSC detector illuminated by the LED placed at a few millimeter of distance from the sensor. Figure \ref{fig:LEDimgLIGHT} (left) shows a single image collected by the DSSC detector illuminated through a paper mask using the LED positioned $\sim 4$ cm of distance from the sensor. These two images shows that the data are correctly transmitted from the readout chain and correctly reordered in the PC receiver.

\subsection{Non-linear sensor characteristic}
As described in \cite{matteo}, the DEPFET sensor \cite{lutz} provides a characteristic response curve, which in the analyzed energy range can be approximated by two linear zones with different slopes with a transition region between them. The first part corresponds to low collected charge and has a higher slope, corresponding to the higher gain needed to achieve single photon sensitivity also at very low photon energies. In the second part, the slope of the curve is lower, corresponding to lower gain at higher signals, needed to extend the dynamic range.



By systematically changing the light intensity it is possible to measure the characteristic response curve of the DSSC system under test to the LED light. In this way, a relative calibration of the pixels is obtained. This can serve as a basis for an absolute calibration by complementing it with the signals measured on the detector when illuminating it with one or more well-known X-ray sources as described in more detail in \cite{georg}.
\begin{figure}[htbp!]
	\centering 
	\includegraphics[width=.45\textwidth,clip]{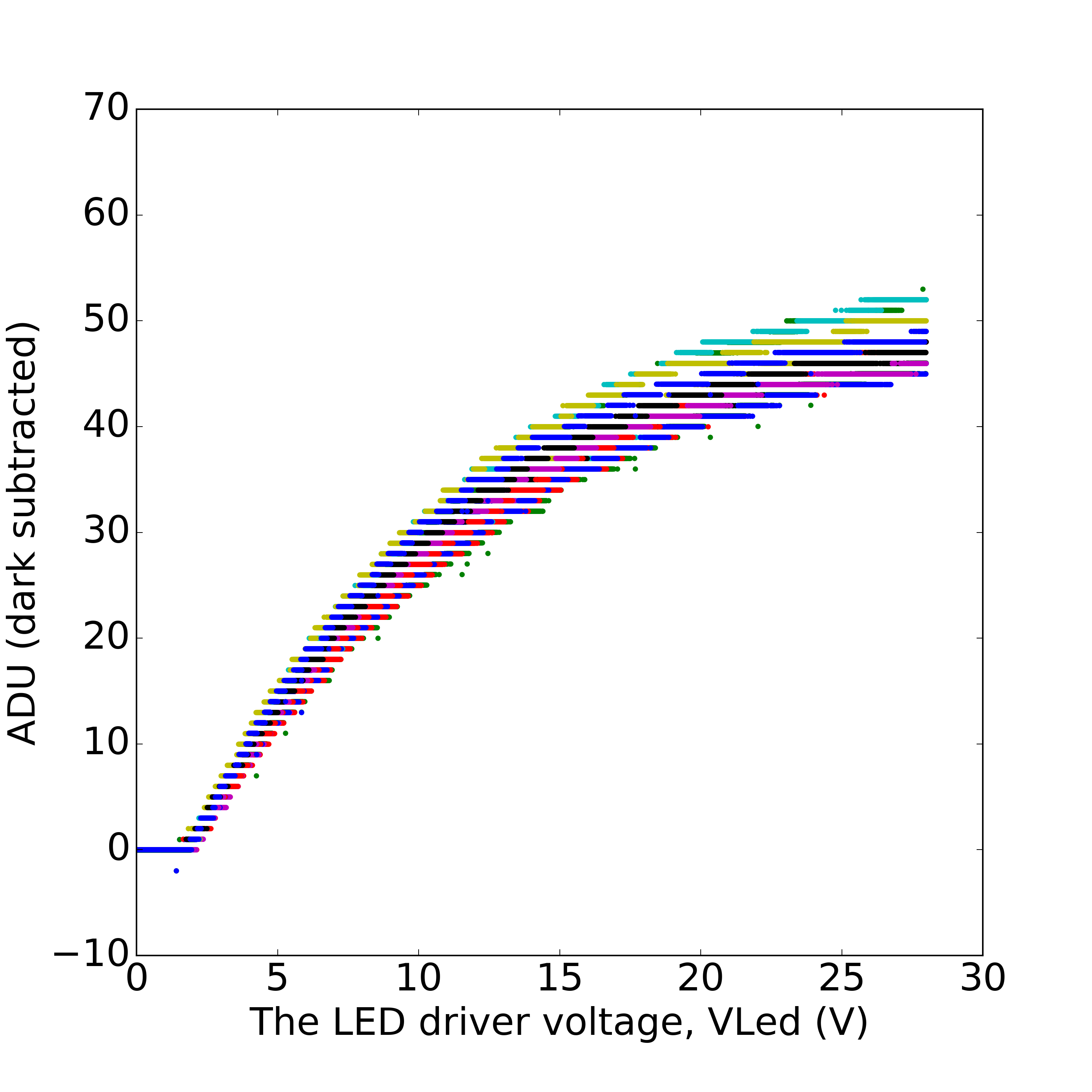}
	\qquad
	\includegraphics[width=.45\textwidth,clip]{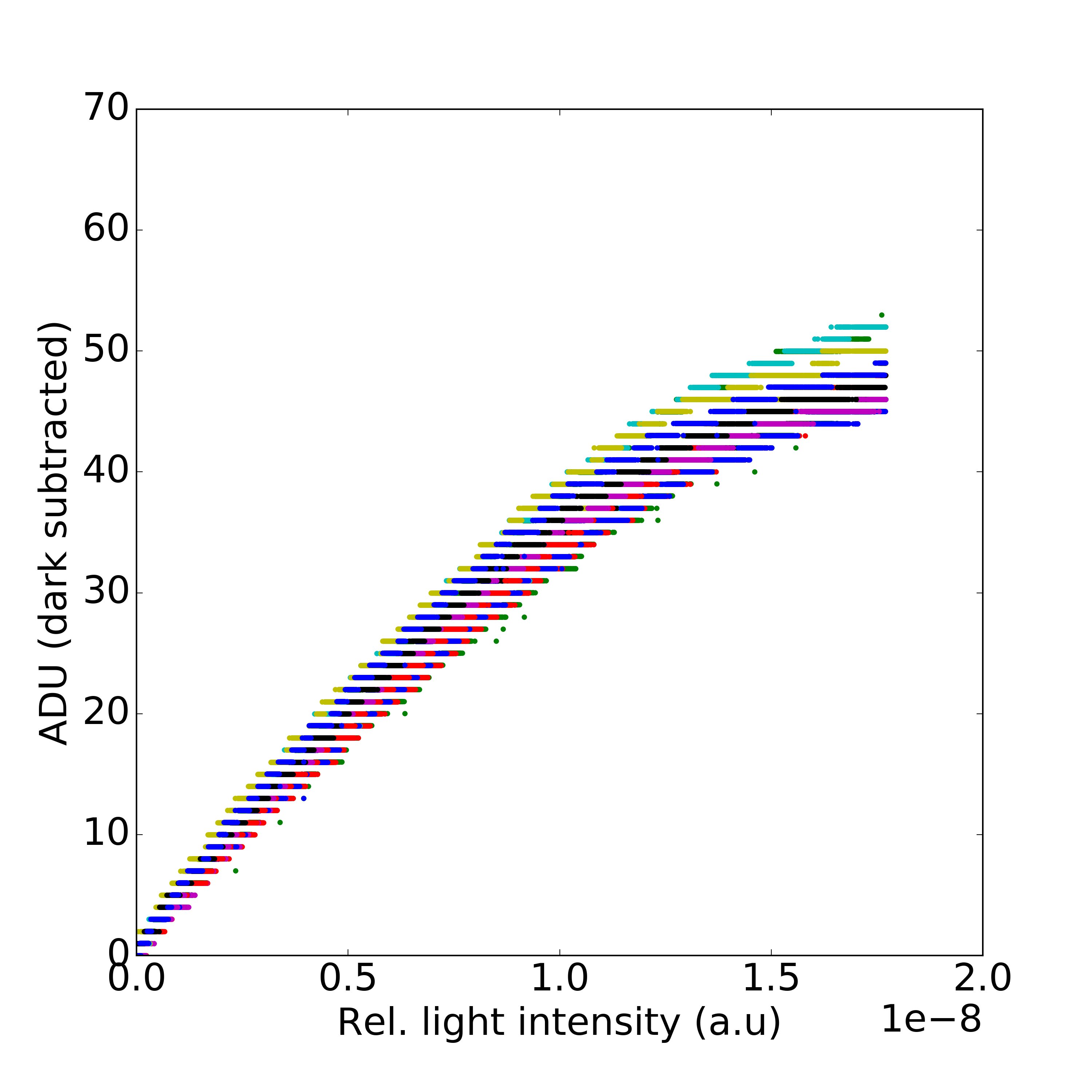}
	\caption{\label{fig:curveadu} Characteristic curve of the mostly illuminated  pixels. Note the different scale on the x-axis.}
\end{figure}
The performed steps are the following: the characteristic relative light intensity as a function of VLed is extracted by using the reference detector as described in section \ref{sec:LEC} and Figure \ref{fig:ledcurve} (right). Then a measurement is performed of the DSSC ADC values as a function of Vled, see Figure \ref{fig:curveadu} (left). Finally,  VLed is converted into relative light intensity using the characteristic light intensity curve previously obtained, Figure \ref{fig:ledcurve} (right); the outcome is shown in Figure \ref{fig:curveadu} (right).

\begin{figure}[htbp!]
	\centering 
	\includegraphics[width=.85\textwidth,clip]{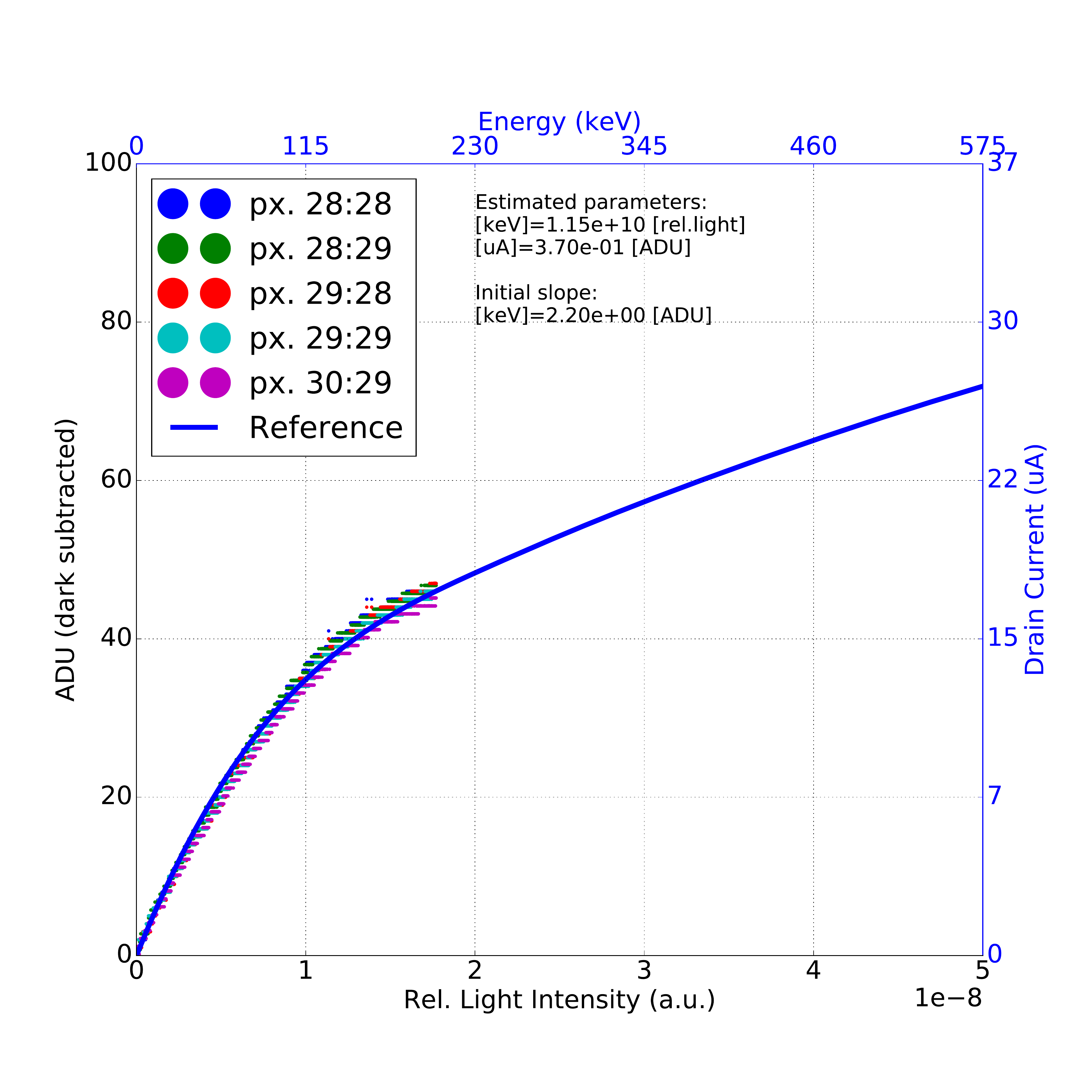}
	\caption{\label{fig:calib} The measured ADC value vs relative light intensity from the LED measurement (points), and the deposited energy vs drain current (blue line), reference curve provided by the sensors producer. The blue axes are refer to the reference curve and the black axes are refer to the measured values.}
\end{figure}

Figure \ref{fig:calib} shows the curve ADC values (dark subtracted) as function of the relative light intensity of a few illuminated neighboring pixels. This curve is superimposed to the expected response curve (drain current vs energy deposited) provided by the sensors producer. In order to superimpose the two curves, a linear transformation has been used. The parameters are shown in Figure \ref{fig:calib}. In this way the relative light intensity corresponds to the estimated deposited energy and the ADC values to the estimated drain current. The curves look in good agreement.



\clearpage
\section{Conclusion}
A 64x64-pixel DSSC ladder detector prototype has been tested with a light source (624 nm) using the complete DSSC readout chain for the first time. The synchronization to the CC has been proven to work correctly. Furthermore, the acquired frames show that the complete readout chain is working properly and the data are correctly reordered. The measured characteristic response curve is promising. Tests of the DSSC ladder with radioactive sources and pulsed X-ray tube are foreseen. Moreover, measurement at synchrotron are already scheduled for the near future. 

%
%
%


\end{document}